
\input phyzzx
\input epsf
\FRONTPAGE
\line{\hfill BROWN-HET-985}
\line{\hfill February 1995}
\bigskip
\titlestyle{NONSINGULAR COSMOLOGY AND PLANCK SCALE PHYSICS
\foot{Invited lecture at the International Workshop on Planck Scale
Physics, Puri, India, Dec. 12-21, 1995. To be published in the proceedings
(World Scientific, Singapore, 1995).}}
\medskip
\author{Robert H. Brandenberger}
\centerline{{\it Department of Physics}}
\centerline{{\it Brown University, Providence, RI 02912, USA}}
\centerline{and}
\centerline{\it Physics Department}
\centerline{\it University of British Columbia}
\centerline{\it Vancouver, B.C. V6T 1Z1, CANADA}
\medskip
\abstract
New Planck scale physics may solve the singularity problems of
classical general relativity and may lead to interesting consequences for
very early Universe cosmology.  Two approaches to these questions are
reviewed in this article.  The first is an effective action approach
to including the effects of Planck scale physics in the basic
framework of general relativity.  It is shown that effective actions
with improved singularity properties can be constructed.  The second
approach is based on superstring theory.  A scenario which eliminates
the big bang singularity and possibly explains the dimensionality of
space-time is reviewed.
\endpage
\chapter{Introduction}

Through its implications for very early Universe cosmology, Planck
scale physics (and specifically string theory) might well
have directly observable consequences for the physical world.  The aim
of this lecture is to explore some possibilities of how this may
occur.

To begin, let us list a couple of problems of modern cosmology: the
homogeneity and flatness problems, the singularity problem, and the
origin of the dimensionality of space-time.

The Universe is observed to be isotropic on large scales to a high
accuracy, the best evidence being the near isotropy of the cosmic
microwave background (CMB).  In standard cosmology there can be no
causal explanation for these observations since the radiation was
emitted from causally disconnected regions of space.  The Universe is
also observed to be approximately spatially flat (the density $\rho$
lies within a factor of 5 from the critical density $\rho_c$ which a
spatially flat Universe would have).  This fact is also a mystery in
the context of standard cosmology, since $\rho = \rho_c$ is an
unstable fixed point in an adiabatically expanding Universe.

Particle physicists have proposed the inflationary Universe
scenario$^1$ as a solution of the homogeneity and flatness problems.
However, closer analysis has shown that standard particle physics
models do not yield inflation$^2$.  Inflation requires a fundamental
scalar field with a reasonably flat potential (in order to have
inflation) and with very small coupling constants (in order that
quantum fluctuations present during inflation do not lead to CMB
temperature anisotropies in excess of those recently detected$^3$;
see Appendix A).  Such potentials are not generic in particle physics
models.

The first challenge from cosmologists to Planck scale physics is
therefore to provide a generic mechanism for inflation.  It may be
that Planck scale physics predicts the type of scalar field potentials
for which successful inflation results. Another possibility is that Planck
scale physics leads to a realization of inflation which does not involve scalar
fields. A possible scenario for this
is suggested in Section 2.  Finally, it may be that Planck scale
physics leads to a solution of the homogeneity and flatness problems
which does not require inflation.

Standard cosmology is plagued by an internal inconsistency.  It
predicts that the Universe started in a ``Big Bang" singularity with
infinite curvature and matter temperature. However, it is known that the
physics on which the standard cosmologyical model is built must break down at
very high temperature and curvature. Therefore, the second challenge for
Planck scale physics is to find a solution of the singularity problem.
Two very different scenarios in which this may happen are suggested in
Sections 2 and 3.

Finally, Planck scale physics (string theory as a concrete example)
allows us to ask questions about the physical world which cannot be
posed in standard physics.  For example, is there a dynamical
mechanism which singles out a Universe in which three space and one time
dimensions are observable?  One mechanism in the context of string
theory will be reviewed in Section 3.

I will review two very different approaches to Planck scale cosmology.
The first is an attempt to incorporate Planck scale effects on the
space-time structure by writing down an effective action for the
space-time metric.  It will be shown that a class of effective actions
exists whose solutions have a less singular structure.  More
specifically, all homogeneous and isotropic solutions are nonsingular
(see Section 2).

In Section 3, I will summarize some aspects of string cosmology and
indicate how in the context of string theory the cosmological
singularities can be avoided.  A dynamical mechanism which explains
why at most three-spatial dimensions are large (and thus observable)
is suggested.

\chapter{A Nonsingular Universe}
\section{Motivation}

Planck scale physics will generate corrections to the Einstein action
which determines the dynamics of the space-time metric $g_{\mu\nu}$.
This can be seen by considering the effective action obtained by
integrating out quantum matter fields in the presence of a dynamical
metric, by calculating first order perturbative quantum gravity
effects, or by studying the low energy effective action of a Planck
scale unified theory such as string theory.

The question we wish to address in this section is whether it is
possible to construct a class of effective actions for gravity which
have improved singularity properties and which predict inflation,
with the constraint that they give the correct low curvature limit.

What follows is a summary of recent work$^{4-7}$ in which we have
constructed an effective action for gravity in which all solutions
with sufficient symmetry are nonsingular.  The theory is a higher
derivative modification of the Einstein action, and is obtained by
a constructive procedure well motivated in analogy with the analysis
of point particle motion in special relativity.  The resulting theory
is asymptotically free in a sense which will be specified below.

A possible objection to our approach is that near a singularity
quantum effects will be important and therefore a classical analysis is
doomed to fail.  This argument is correct in the usual picture in
which at high curvatures there are large fluctuations and space-time
becomes more like a ``quantum foam."  However, in our theory, at high
curvature space-time becomes highly regular and thus a classical
analysis of space-time is self-consistent.  The property of asymptotic
freedom is essential in order to reach this conclusion.

Our aim is to construct a theory with the property that the metric
$g_{\mu\nu}$ approaches the de Sitter metric $g_{\mu\nu}^{DS}$, a
metric with maximal symmetry which admits a geodesically complete and
nonsingular extension, as the curvature $R$ approaches the Planck
value $R_{pl}$.  Here, $R$ stands for any curvature invariant.
Naturally, from our classical considerations, $R_{pl}$ is a free
parameter.  However, if our theory is connected with Planck scale
physics, we expect $R_{pl}$ to be set by the Planck scale.

\epsfxsize=6in \epsfbox{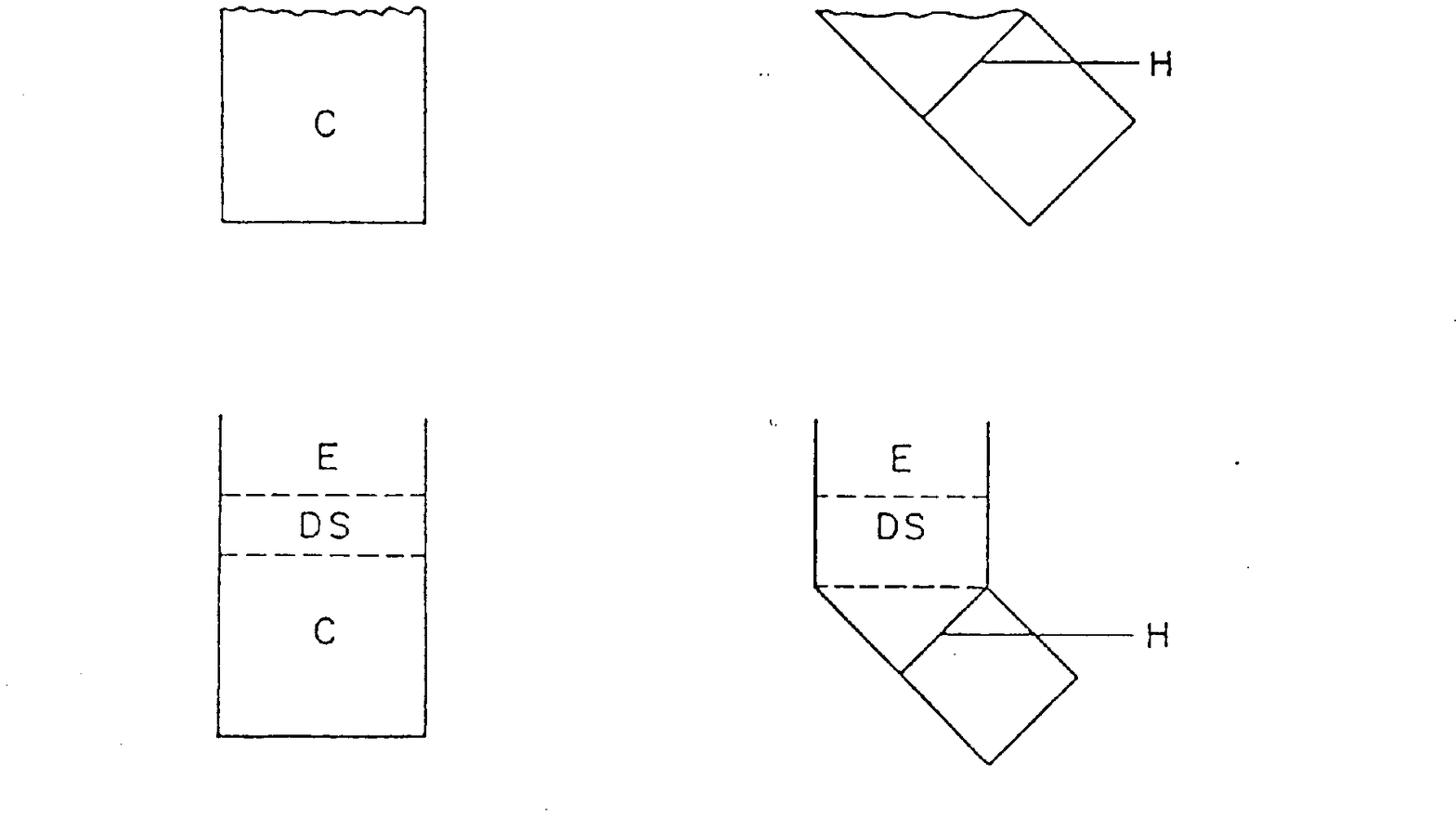}
{\baselineskip=13pt
\noindent{\bf Figure 1:} Penrose diagrams for collapsing Universe (left) and
black hole (right) in Einstein's theory (top) and in the nonsingular Universe
(bottom). C, E, DS and H stand for contracting phase, expanding phase, de
Sitter phase and horizon, respectively, and wavy lines indicate singularities.}

If successful, the above construction will have some very appealing
consequences.  Consider, for example, a collapsing spatially
homogeneous Universe.  According to Einstein's theory, this Universe
will collapse in finite proper time to a final ``big crunch" singularity (top
left Penrose diagram of Figure 1).
In our theory, however, the Universe will approach a de Sitter model as
the curvature increases.  If the
Universe is closed, there will be a de Sitter bounce followed by
re-expansion (bottom left Penrose diagram in Figure 1).  Similarly, in our
theory spherically
symmetric vacuum  solutions would be nonsingular, i.e., black holes
would have no singularities in their centers.  The structure of a
large black hole would be unchanged compared to what is predicted by
Einstein's theory (top right, Figure 1) outside and even slightly inside the
horizon, since
all curvature
invariants are small in those regions.  However, for $r \rightarrow 0$
(where $r$ is the radial Schwarzschild coordinate), the solution
changes and approaches a de Sitter solution (bottom right, Figure 1).  This
would have interesting consequences for the black hole information
loss problem.

To motivate our effective action construction, we turn to a well known
analogy, point particle motion in the theory of special relativity.

\section{An Analogy}

 The transition from the Newtonian theory of point particle motion to
the special relativistic theory transforms a theory with no bound on
the velocity into one in which there is a limiting velocity, the speed
of light $c$ (in the following we use units in which $\hbar = c = 1$).
This transition can be obtained$^4$ by starting with the action of a
point particle with world line $x(t)$:
$$
S_{\rm old} = \int dt {1\over 2} \dot x^2 \, , \eqno\eq
$$
and adding$\varphi^8$  a Lagrange multiplier which couples to $\dot
x^2$, the quantity to be made finite, and which has a potential
$V(\varphi)$:
$$
S_{\rm new} = \int dt \left[ {1\over 2} \dot x^2 + \varphi \dot x^2 -
V (\varphi) \right] \, .\eqno\eq
$$
{}From the constraint equation
$$
\dot x^2 = {\partial V\over{\partial \varphi}} \, , \eqno\eq
$$
it follows that $\dot x^2$ is limited provided $V(\varphi)$ increases
no faster than linearly in $\varphi$ for large $|\varphi|$.  The small
$\varphi$ asymptotics of $V(\varphi)$ is determined by demanding that
at low velocities the correct Newtonian limit results:
$$
\eqalign{V (\varphi) \sim \varphi^2 \> & {\rm as} \> |\varphi|
\rightarrow 0 \, , \cr
V (\varphi) \sim \varphi \> & {\rm as} \> |\varphi| \rightarrow \infty
\, . } \eqno\eq
$$
Choosing the simple interpolating potential
$$
V (\varphi) = {2 \varphi^2\over{1 + 2 \varphi}} \, , \eqno\eq
$$
the Lagrange multiplier can be integrated out, resulting in the well-known
action
$$
S_{\rm new} = {1\over 2} \int dt \sqrt{1 - \dot x^2} \eqno\eq
$$
for point particle motion in special relativity.

\section{Construction}

 Our procedure for obtaining a nonsingular Universe theory$^4$ is based
on generalizing the above Lagrange multiplier construction to gravity.
Starting from the Einstein action, we can introduce a Lagrange
multiplier $\varphi_1$ coupled to the Ricci scalar $R$ to obtain a
theory with limited $R$:
$$
S = \int d^4 x \sqrt{-g} (R + \varphi_1 \, R + V_1 (\varphi_1) ) \, ,
\eqno\eq
$$
where the potential $V_1 (\varphi_1)$ satisfies the asymptotic
conditions (2.4).

However, this action is insufficient to obtain a nonsingular gravity
theory.  For example, singular solutions of the Einstein equations
with $R=0$ are not effected at all.  The minimal requirements for a
nonsingular theory is that \underbar{all} curvature invariants remain
bounded and the space-time manifold is geodesically complete.
Implementing the limiting curvature hypothesis$^9$, these conditions
can be reduced to more manageable ones.  First, we choose one
curvature invariant $I_1 (g_{\mu\nu})$ and demand that it be
explicitely bounded, i.e., $|I_1| < I_1^{pl}$, where $I_1^{pl}$ is the
Planck scale value of $I_1$.  In a second step, we demand that as $I_1
(g_{\mu\nu})$ approaches $I_1^{pl}$, the metric $g_{\mu\nu}$ approach
the de Sitter metric $g^{DS}_{\mu\nu}$, a definite nonsingular metric
with maximal symmetry.  In this case, all curvature invariants are
automatically bounded (they approach their de Sitter values), and the
space-time can be extended to be geodesically complete.

Our approach is to implement the second step of the above procedure by
another Lagrange multiplier construction$^4$.  We look for a curvature
invariant $I_2 (g_{\mu\nu})$ with the property that
$$
I_2 (g_{\mu\nu}) = 0 \>\> \Leftrightarrow \>\> g_{\mu\nu} =
g^{DS}_{\mu\nu} \, , \eqno\eq
$$
introduce a second Lagrange multiplier field $\varphi_2$ which couples
to $I_2$ and choose a potential $V_2 (\varphi_2)$ which forces $I_2$
to zero at large $|\varphi_2|$:
$$
S = \int d^4  x \sqrt{-g} [ R + \varphi_1 I_1 + V_1 (\varphi_1) +
\varphi_2 I_2 + V_2 (\varphi_2) ] \, , \eqno\eq
$$
with asymptotic conditions (2.4) for $V_1 (\varphi_1)$ and conditions
$$
\eqalign{V_2 (\varphi_2) & \sim {\rm const} \>\> {\rm as} \> |
\varphi_2 | \rightarrow \infty \cr
V_2 (\varphi_2) & \sim \varphi^2_2 \>\> {\rm as} \> |\varphi_2 |
\rightarrow 0 \, ,} \eqno\eq
$$
for $V_2 (\varphi_2)$.  The first constraint forces $I_2$ to zero, the
second is required in order to obtain the correct low curvature limit.

These general conditions are reasonable, but not sufficient in order
to obtain a nonsingular theory.  It must still be shown that all
solutions are well behaved, i.e., that they asymptotically reach the
regions $|\varphi_2| \rightarrow \infty$ of phase space (or that
they can be controlled in some other way).  This must be done for a
specific realization of the above general construction.

\section{Specific Model}

At the moment we are only able to find an invariant $I_2$ which
singles out de Sitter space by demanding $I_2 = 0$ provided we assume
that the metric has special symmetries.  The choice
$$
I_2 = (4  R_{\mu\nu} R^{\mu\nu} - R^2 + C^2)^{1/2} \, , \eqno\eq
$$
singles out the de Sitter metric among all homogeneous and isotropic
metrics (in which case adding $C^2$, the Weyl tensor square, is
superfluous), all homogeneous and anisotropic metrics, and all
radially symmetric metrics.

We choose the action$^{4,5}$
$$
S = \int d^4 x \sqrt{-g} \left[ R + \varphi_1 R - (\varphi_2 +
{3\over{\sqrt{2}}} \varphi_1) I_2^{1/2} + V_1 (\varphi_1) + V_2
(\varphi_2) \right] \eqno\eq
$$
with
$$
V_1 (\varphi_1) = 12 \, H^2_0 {\varphi^2_1\over{1 + \varphi_1}} \left( 1
- {\ln (1 + \varphi_1)\over{1 + \varphi_1}} \right) \eqno\eq
$$
$$
V_2 (\varphi_2) = - 2 \sqrt{3} \, H^2_0 \, {\varphi^2_2\over{1 +
\varphi^2_2}} \, . \eqno\eq
$$

The general equations of motion resulting from this action are quite
messy.  However, when restricted to homogeneous and isotropic metrics
of the form
$$
ds^2 = dt^2 - a (t)^2 (dx^2 + dy^2 + dz^2) \, , \eqno\eq
$$
the equations are fairly simple.  With $H = \dot a / a$, the two
$\varphi_1$ and $\varphi_2$ constraint equations are
$$
H^2 = {1\over{12}} V^\prime_1 \eqno\eq
$$
$$
\dot H = - {1\over{2\sqrt{3} }} V^\prime_2 \, , \eqno\eq
$$
and the dynamical $g_{00}$ equation becomes
$$
3 (1 - 2 \varphi_1) H^2 + {1\over 2} (V_1 + V_2) = \sqrt{3} H (\dot
\varphi_2 + 3 H \varphi_2) \, . \eqno\eq
$$
The phase space of all vacuum configurations is the half plane $\{
(\varphi_1 \geq 0, \, \varphi_2) \}$.  Equations (2.16) and (2.17)
can be used to express $H$ and $\dot H$ in terms of $\varphi_1$ and
$\varphi_2$.  The remaining dynamical equation (2.18) can then be recast
as
$$
{d \varphi_2\over{d \varphi_1}} = - {V_1^{\prime\prime}\over{4
V^\prime_2}} \, \left[ - \sqrt{3} \varphi_2 + (1 - 2\varphi_1) -
{2\over{V^\prime_1}} (V_1 + V_2) \right] \, . \eqno\eq
$$
The solutions can be studied analytically in the asymptotic regions
and numerically throughout the entire phase space.

The resulting phase diagram of vacuum solutions is sketched in Fig. 2
(for numerical results, see Ref. 5).  The point $(\varphi_1, \,
\varphi_2) = (0,0)$ corresponds to Minkowski space-time $M^4$, the
regions $|\varphi_2 | \rightarrow \infty$ to de Sitter space.  As
shown, all solutions either are periodic about $M^4$ or else they
asymptotically approach de Sitter space.  Hence, all solutions are
nonsingular.  This conclusion remains unchanged if we add spatial
curvature to the model.

\epsfxsize=6in \epsfbox{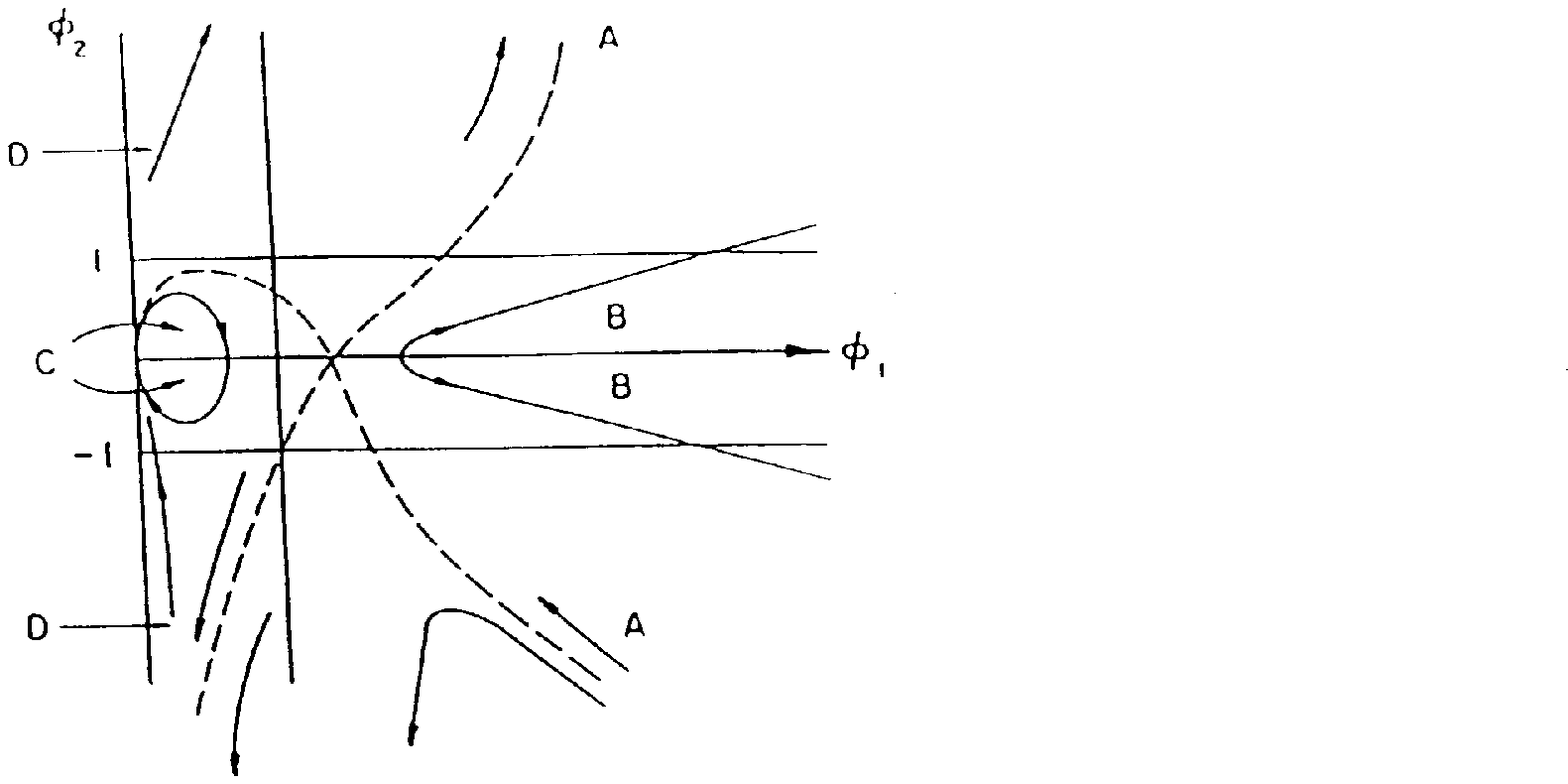}
{\baselineskip=13pt
\noindent{\bf Figure 2:} Phase diagram of the homogeneous and isotropic
solutions of the nonsingular Universe. The asymptotic regions are labelled by
A, B, C and D, flow lines are indicated by arrows.}

One of the most interesting properties of our theory is asymptotic
freedom$^5$, i.e., the coupling between matter and gravity goes to
zero at high curvatures.  It is easy to add matter (e.g., dust or
radiation) to our model by taking the combined action
$$
S = S_g + S_m \, , \eqno\eq
$$
where $S_g$ is the gravity action previously discussed, and $S_m$ is
the usual matter action in an external background space-time metric.

We find$^5$ that in the asymptotic de Sitter regions, the trajectories of
the solutions in the $(\varphi_1, \, \varphi_2)$ plane are unchanged
by adding matter.  This applies, for example, in a phase of de Sitter
contraction when the matter energy density is increasing exponentially
but does not affect the metric.  The physical reason for asymptotic
freedom is obvious: in the asymptotic regions of phase space, the
space-time curvature approaches its maximal value and thus cannot be
changed even by adding an arbitrary high matter energy density.

Naturally, the phase space trajectories near $(\varphi_1, \,
\varphi_2) = (0,0)$ are strongly effected by adding matter.  In
particular, $M^4$ ceases to be a stable fixed point of the evolution
equations.

\section{Connection with Dilaton Gravity}

 The low energy effective actions for the space-time metric in 4
dimensions which come from string theory are only known
perturbatively.  They contain higher derivative terms, but not if the
exact same form as the ones used in our construction.  The connection
between our limiting curvature construction and string theory-motivated
effective actions is more apparent in two
 space-time dimensions$^{6,7}$.

The most general renormalizable Lagrangian for string-induced dilaton
gravity is
$$
{\cal L} = \sqrt{-g} [ D(\varphi) R + G (\varphi) (\nabla \varphi)^2 +
H (\varphi) ] \, , \eqno\eq
$$
where $\varphi (x,t)$ is the dilaton.  In two space-time dimensions,
the kinetic term for $\varphi$ can be eliminated, resulting in a
Lagrangian (in terms of rescaled fields) of the form
$$
{\cal L} = \sqrt{-g} [ D(\varphi) R + V (\varphi) ] \, . \eqno\eq
$$

We can now apply the limiting curvature construction to find classes
of potentials for which the theory has nonsingular black hole$^6$ and
cosmological$^7$ solutions.  In the following, we discuss the
nonsingular two-dimensional black hole.

To simplify the algebra, the dilaton is redefined such that
$$
D (\varphi) = {1\over \varphi} \, . \eqno\eq
$$
The most general static metric can be written as
$$
ds^2 = f (r) dt^2 - g (r) dr^2 \eqno\eq
$$
and the gauge choice
$$
g (r) = f (r)^{-1} \eqno\eq
$$
is always possible.  The variational equations are
$$
f^\prime = - V (\varphi) {\varphi^2\over \varphi^\prime} \, , \eqno\eq
$$
$$
\left( {\varphi^\prime\over \varphi^2} \right)^\prime = 0 \eqno\eq
$$
and
$$
\varphi^{-2} R = {\partial V\over{\partial \varphi}} \, , \eqno\eq
$$
where a prime denotes the derivative with respect to $r$.

Equation (2.27) can be integrated to find (after rescaling $r$)
$$
\varphi = {1\over{Ar}} \, . \eqno\eq
$$
To give the correct large $r$ behavior for the metric, we need to
impose that
$$
f (r) \rightarrow 1 - {2m\over r} \>\>\> {\rm as} \> r \rightarrow
\infty \, . \eqno\eq
$$
{}From (2.26) this leads to the asymptotic condition
$$
V (\varphi) \rightarrow 2 m A^3 \varphi^2 \>\>\> {\rm as} \> \varphi
\rightarrow 0 \, . \eqno\eq
$$
The limiting curvature hypothesis requires that $R$ be bounded as
$\varphi \rightarrow \infty$.  From (2.28) this implies
$$
V (\varphi) \rightarrow {2\over{\ell^2 \varphi}} \>\>\> {\rm as} \>
\varphi \rightarrow \infty \, , \eqno\eq
$$
where $\ell$ is a constant which determines the limiting curvature.
As an interpolating potential we can choose
$$
V (\varphi) = {2 m A^3 \varphi^2\over{1+ m A^3 \ell^2 \varphi^3}} \, ,
\eqno\eq
$$
which allows (2.26) to be integrated explicitly$^6$ to obtain $f(r)$.

The resulting metric coefficient $f(r)$ describes a nonsingular black
hole with a single horizon at $r \simeq 2m$.  The metric is
indistinguishable from the usual Schwarzschild metric until far inside
of the horizon, where our $f(r)$ remains regular and obtains vanishing
derivative at $r = 0$, which allows for a geodesically complete
extension of the manifold.

\section{Discussion}

We have shown that a class of higher derivative extensions of the
Einstein theory exist for which many interesting solutions are
nonsingular.  Our class of models is very special.  Most higher
derivative theories of gravity have, in fact, much worse singularity
properties than the Einstein theory.  What is special about our class
of theories is that they are obtained using a well motivated Lagrange
multiplier construction which implements the limiting curvature
hypothesis.  We have shown that
\item{\rm i)} all homogeneous and isotropic solutions are
nonsingular$^{4,5}$
\item{\rm ii)} the two-dimensional black holes are nonsingular$^6$
\item{\rm iii)} nonsingular two-dimensional cosmologies exist$^7$.

\noindent
We also have evidence that four-dimensional black holes and
anisotropic homogeneous cosmologies are nonsingular$^{10}$.

By construction, all solutions are de Sitter at high curvature.  Thus,
the theories automatically have a period of inflation (driven by the
gravity sector in analogy to Starobinsky inflation$^{11}$) in the
early Universe.

A very important property of our theories is asymptotic freedom.  This
means that the coupling between matter and gravity goes to zero at
high curvature, and might lead to an automatic suppression mechanism
for scalar fluctuations.

In two space-time dimensions, there is a close connection between
dilaton gravity and our construction.  In four dimensions, the
connection between fundamental physics and our class of effective
actions remains to be explored.

\chapter{Aspects of String Cosmology}
\section{Motivation}

In the previous section we studied effective actions for the space-time
metric which might arise in the intermediate energy regime of a fundamental
theory such as string theory.  However, it is also of interest to
explore the predictions of string theory which depend specifically on
the ``stringy" aspects of the theory and which are lost in any field
theory limit.  It is to a description of a few of the string-specific
cosmological aspects to which we turn in this section.

\section{Implications of Target Space Duality}
Target space duality$^{12}$ is a symmetry specific to string theory.
As a simple example, consider a superstring background in which all
spatial dimensions are toroidally compactified with equal radii.  Let
$R$ denote the radius of the torus.

The spectrum of string states is spanned by oscillatory modes which
have energies independent of $R$, by momentum modes whose energies
$E_n$ (with integer $n$) are
$$
E_n = {n\over R} \, , \eqno\eq
$$
and by winding modes with energies $E^\prime_m$ ($m$ integer)
$$
E^\prime_m = mR \, . \eqno\eq
$$

Target space duality is a symmetry between two superstring theories,
one on a background with radius $R$, the other on a background of
radius $1/R$, under which winding and momentum modes are interchanged.

Target space duality has interesting consequences for string
cosmology$^{13}$.  Consider a background with adiabatically changing
$R(t)$.  While $R(t) \gg 1$, most of the energy in thermal equilibrium
resides in the momentum modes.  The position eigenstates $|x >$ are
defined as in quantum field theory in terms of the Fourier transform
of the momentum eigenstates $|p >$
$$
|x > = \sum\limits_p e^{i x \cdot p} |p > \, . \eqno\eq
$$
However, for $R (t) \ll 1$, most of the energy flows into winding
modes, and it takes much less energy to measure the ``dual distance"
$| \tilde x >$ than $|x >$, where
$$
| \tilde x > = \sum\limits_w e^{i \tilde x \cdot w} | w > \eqno\eq
$$
is defined in terms of the winding modes $| w>$.

We conclude that target space duality in string theory leads to a
minimum physical length in string cosmology.  As $R(t)$ decreases
below 1, the measured length starts to increase again.  This could
lead to a bouncing or oscillating cosmology$^{13}$.

It is  well known that for strings in thermal equilibrium there is a
maximal temperature, the Hagedorn temperature$^{14}$.  Target space
duality implies that in thermal equilibrium the temperature in an
adiabatically varying string background begins to decrease once $R(t)$
falls below 1:
$$
T \left({1\over R} \right) = T(R) \, . \eqno\eq
$$
Thus, the $T(R)$ curve in string cosmology is nonsingular and very
different from its behavior in standard cosmology.  For further
discussions of the thermodynamics of strings see, e.g., Refs. 15 and
16 and references therein.

\section{Strings and Space-Time Dimensionality}

Computations$^{13}$ using the microcanonical ensemble show that for
all spatial directions compactified at large total energy $E$, the
entropy $S$ is proportional to $E$:
$$
S = \beta_H E \, , \eqno\eq
$$
with $\beta_H$ denoting the inverse of the Hagedorn temperature $T_H$.
Thus, the $E(R)$ curve in string cosmology is very different from the
corresponding curve in standard cosmology.

For large $R \gg 1$, most of the energy in a gas of strings in thermal
equilibrium will flow into momentum modes, and the thermodynamics will
approach that of an ideal gas of radiation for which
$$
E (R) \sim {1\over R} \, . \eqno\eq
$$
By duality, for small $R$
$$
E (R) \sim R \, . \eqno\eq
$$

If, however, for some reason the string gas falls out of equilibrium,
the $E(R)$ curve will look very different.  Starting at $R= 1$ with a
temperature approximately equal to $T_H$, a large fraction of the
energy will reside in winding modes.  If these winding modes cannot
annihilate, thermal equilibrium will be lost, and the energy in
winding modes will increase linearly in $R$, and thus for large $R$:
$$
E (R) \sim R \, . \eqno\eq
$$

Newtonian intuition tells us that out of equilibrium winding modes
with an energy relation (3.9) will prevent the background space from
expanding$^{13}$.  The equation of state corresponding to a gas of
straight strings is
$$
p = - {1\over N} \rho \eqno\eq
$$
where $p$ and $\rho$ denote pressure and energy density, respectively, and
$N$ is the number of spatial dimensions.
According to standard general relativity, an equation of state with
negative pressure will lead to more rapid expansion of the background.
It turns out that the Newtonian intuition is the correct one and that
general relativity gives the wrong answer$^{17}$.  At high densities,
the specific stringy effects -- in particular target space duality
 -- become crucial.

The Einstein action violates duality.  In order to restore duality, it
is necessary to include the dilaton in the effective action for the
string background.  The action for dilaton gravity is
$$
S = \int d^{N+1} x \sqrt{-g} e^{-2 \phi} [ R+ 4 (D \phi)^2 ] \eqno\eq
$$
where $\phi$ is the
dilaton.  It is convenient to use new fields $\varphi$ and $\lambda$
defined by
$$
a (t) = e^{\lambda t} \eqno\eq
$$
and
$$
\varphi = 2 \phi - N \lambda \, . \eqno\eq
$$
The action (3.11) has the duality symmetry
$$
\lambda \rightarrow - \lambda, \> \varphi \rightarrow \varphi \, .
\eqno\eq
$$

The variational equations of motion derived from (3.11) for a
homogeneous and isotropic model are$^{17, 18}$
$$
\eqalign{& \dot \varphi^2 = e^\varphi E + N \dot \lambda^2 \cr
& \ddot \lambda - \dot \varphi \dot \lambda = {1\over 2} e^\varphi P \cr
& \ddot \varphi = {1\over 2} e^\varphi E + N \dot \lambda^2 \, ,
}\eqno\eq
$$
where $P$ and $E$ are total pressure and energy, respectively.  For a
winding mode-dominated equation of state (and neglecting friction
terms) the equation of motion for $\lambda (t)$ becomes
$$
\ddot \lambda = - {1\over{2N}} e^\varphi E(\lambda) \, , \eqno\eq
$$
which corresponds to motion in a confining potential.  Hence, winding modes
prevent the background toroidal
dimensions from expanding.

These considerations may be used to put forward the conjecture$^{13}$
that string cosmology will single out three as the maximum number of
spatial dimensions which can be large ($R \gg 1$ in Planck units).
The argument proceeds as follows.  Space can, starting from an initial
state with $R \sim 1$ in all directions, only expand if thermal
equilibrium is maintained, which in turn is only possible if the
winding modes can annihilate.  This can only happen in at most three spatial
 dimensions (in a higher number the probability for
intersection of the world sheets of two strings is zero).  In the
critical dimension for strings, $N=3$, the evolution of a string gas
has been studied extensively in the context of the cosmic string
theory (see e.g., Refs. 19 and 20 for recent reviews).  The winding
modes do, indeed, annihilate, leaving behind a string network with
about one winding mode passing through each Hubble volume.  Thus, in
string cosmology only three spatial dimensions will become large
whereas the others will be confined to Planck size by winding modes.

\chapter{Conclusions}

Planck scale physics may have many observational consequences and may
help cosmologists solve some of the deep puzzles concerning the origin
of inflation, the absence of space-time singularities and the
dimensionality of space-time.

A lot of work needs to be done before these issues are properly
understood.  I have outlined two ways to address some of these
questions.  The first investigation was based on classical physics
and attempted
to analyze what can be said about the origin of inflation and about
singularities from an effective action approach to gravity.  We
constructed a class of higher derivative gravity actions without
singular cosmological solutions (i.e., no singular homogeneous and
isotropic solutions) and which automatically give rise to inflation.

The second approach was an exploration of some of the cosmological
consequences of target space duality in string theory.  A nonsingular
cosmological scenario was proposed which might even explain why only
three-spatial dimensions are large.

\ack

My sincere thanks goes to Professor Jnan Maharana and the other
members of the Institute of Physics in Bhubaneswar, for organizing an
extremely stimulating workshop and for their warm hospitality.  I wish
to thank my collaborators Richhild Moessner, Masoud Mohazzab, Slava
Mukhanov, Andrew Sornborger, Mark Trodden and Cumrun Vafa for the joy
of collaboration.  I also acknowledge stimulating discussions with
Sanjay Jain and Ajit Srivastava at the meeting.  I am grateful to Bill
Unruh for his generous hospitality at UBC where this article was
written up.

This work is supported in part by the US Department of Energy under
Grant DE-FG0291ER40688, Task A (Brown) and be the Canadian NSERC under
Grant 580441.  Travel support from ICTP Trieste is gratefully
acknowledged.

\Appendix{A}
\noindent
\underbar{\bf Conditions for Successful Inflation}

There are two sets of constraints on scalar field-driven inflationary
Universe models.  Firstly, the equation of state of the scalar field
must be compatible with inflation.  Secondly, the induced quantum
fluctuations must be consistent with the limits from recent CMB
anisotropy experiments.

In order for the equation of state of a scalar field $\varphi$
(assumed to be spatially homogeneous for simplicity) to be consistent
with inflation, it is sufficient to require that the energy density be
dominated by the potential $V(\varphi)$ and that $\varphi$ is slowly
rolling, i.e., that the acceleration term $\ddot \varphi$ in the
equation of motion
$$
\ddot \varphi + 3 H \dot \varphi = - V^\prime (\varphi) \eqno\eq
$$
is negligible.  In the above, $H$ is the Hubble expansion rate and a
prime denotes the derivative with respect to $\varphi$.

The condition for the energy density $\rho$ to be potential dominated
is that
$$
\dot \varphi^2 \ll V (\varphi) \, . \eqno\eq
$$
In order for this condition to be maintained over many Hubble
expansion times $\Delta t = H^{-1}$, it is sufficient to require that
$\ddot \varphi$ in (A.1) is negligible.  A necessary condition for this
to be the case is
$$
\big| {V^{\prime\prime}\over{24 \pi GV}} \big| \ll 1 \, , \eqno\eq
$$
$G$ being Newton's constant.  This condition is obtained by taking
Equation (A.1) without the $\ddot \varphi$ term (i.e., in the slow
rolling approximation), solving for $\dot \varphi$, differentiating to
obtain $\ddot \varphi$, and requiring that the result be smaller in
absolute magnitude than $V^\prime (\varphi)$.  The validity of (A.2) is
assumed in the derivation.

Quantum fluctuations produced during inflation generate energy density
fluctuations on scales relevant to cosmology (see e.g., Ref. 21
for a recent review).  In particular, CMB anisotropies are induced.
In order for these anisotropies not to exceed the recent observational
results$^{21}$, the following condition on the
scalar field potential must be satisfied$^{3}$
$$
{V (\varphi_i)\over{(\Delta \varphi)^4}} < 10^{-6} \, . \eqno\eq
$$
Here, $\varphi_i$ is the value of the inflaton field at the time when
perturbations on present day Hubble radius scale are produced, and
$\Delta \varphi$ is the change in $\varphi$ between $\varphi_i$ and
the value of $\varphi$ at the end of inflation, at which point it is
assumed that the potential $V$ vanishes (see Figure 3).

In the context of standard particle physics, it is hard to satisfy all
three conditions (A.2), (A.3) and, in particular, (A.4).  For example, in
a model of chaotic inflation with potential $V(\varphi) = \lambda \varphi^4$,
the condition (A.4) implies
$$
\lambda < 10^{-8} \, . \eqno\eq
$$

\epsfxsize=5.5in \epsfbox{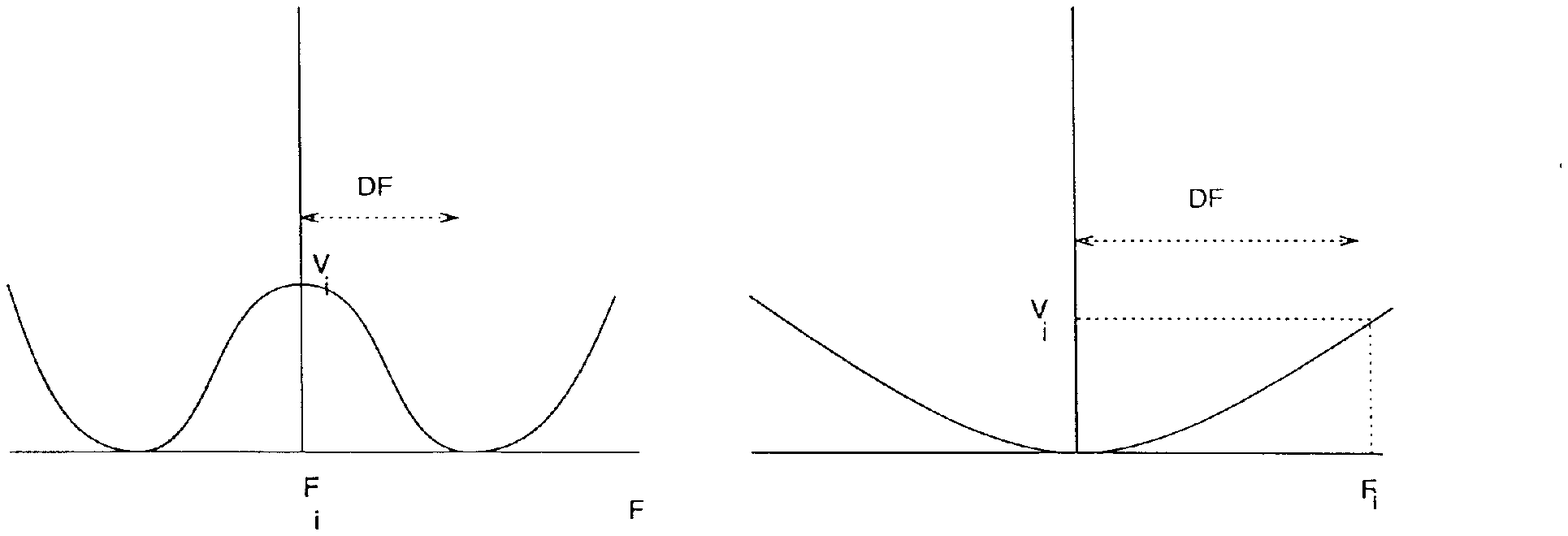}
{\baselineskip=13pt
\noindent{\bf Figure 3:} The potential energy density $V(\varphi)$ in models of
new inflation (left) and chaotic inflation (right). $\varphi$ is denoted by
$F$, and $\Delta \varphi$ by $DF$.}

If the correct resolution of the homogeneity and flatness problems is
based on some fundamental principle rather than on special choices of
parameters in some rather ad hoc model, then new physics seems to be required.

\REF\one{A. Guth, {\it Phys. Rev.} {\bf D23}, 247 (1981).}
\REF\two{A. Linde, {\it Phys. Lett.} {\bf B129}, 177 (1983); \nextline
G. Mazenko, W. Unruh and R. Wald, {\it Phys. Rev.} {\bf D31}, 273
(1985).}
\REF\three{F. Adams, K. Freese and A. Guth, {\it Phys. Rev.} {\bf D43}, 965
(1991).}
\REF\four{V. Mukhanov and R. Brandenberger, {\it Phys. Rev. Lett.} {\bf
68}, 1969 (1992).}
\REF\five{R. Brandenberger, V. Mukhanov and A. Sornborger, {\it Phys.
Rev.} {\bf D48}, 1629 (1993).}
\REF\six{M. Trodden, V. Mukhanov and R. Brandenberger, {\it Phys.
Lett.} {\bf B316}, 483 (1993).}
\REF\seven{R. Moessner and M. Trodden, `Singularity-Free Two
Dimensional Cosmologies,' Brown preprint BROWN-HET-942, {\it Phys.
Rev. D}, in press (1995).}
\REF\eight{B. Altshuler, {\it Class. Quant. Grav.} {\bf 7}, 189
(1990).}
\REF\nine{M. Markov, {\it Pis'ma Zh. Eksp. Theor. Fiz.} {\bf 36}, 214
(1982); \nextline
M. Markov, {\it Pis'ma Zh. Eksp. Theor. Fiz.} {\bf 46}, 342 (1987);
\nextline
V. Ginsburg, V. Mukhanov and V. Frolov,  {\it Pis'ma Zh. Eksp. Theor. Fiz.}
{\bf 94}, 3 (1988); \nextline
V. Frolov, M. Markov and V. Mukhanov, {\it Phys. Rev.} {\bf D41}, 383
(1990).}
\REF\ten{R. Brandenberger, M. Mohazzab, V. Mukhanov, A. Sornborger and
M. Trodden, in preparation (1995).}
\REF\eleven{A. Starobinsky, {\it Phys. Lett.} {\bf B91}, 99 (1980).}
\REF\twelve{K. Kikkawa and M. Yamasaki, {\it Phys. Lett.} {\bf B149},
357 (1984); \nextline
N. Sakai and I. Senda, {\it Prog. Theor. Phys.} {\bf 75}, 692 (1986);
\nextline
B. Sathiapalan, {\it Phys. Rev. Lett.} {\bf 58}, 1597 (1987);
\nextline
P. Ginsparg and C. Vafa, {\it Nucl. Phys.} {\bf B289}, 414 (1987).}
\REF\thirteen{R. Brandenberger and C. Vafa, {\it Nucl. Phys.} {\bf
B316}, 391 (1989).}
\REF\fourteen{R. Hagedorn, {\it Nuovo Cimento Suppl.} {\bf 3}, 147
(1965).}
\REF\fifteen{D. Mitchell and N. Turok, {\it Nucl. Phys.} {\bf B294},
1138 (1987).}
\REF\sixteen{N. Deo, S. Jain and C.-I. Tan, {\it Phys. Rev.} {\bf
D40}, 2626 (1989).}
\REF\seventeen{A. Tseytlin and C. Vafa, {\it Nucl. Phys.} {\bf B372},
443 (1992).}
\REF\eighteen{G. Veneziano, {\it Phys. Lett.} {\bf B265}, 287 (1991).}
\REF\nineteen{R. Brandenberger, {\it Int. J. Mod. Phys.} {\bf A9},
2117 (1994).}
\REF\twenty{A. Vilenkin and E.P.S. Shellard, `Cosmic Strings and
other Topological Defects' (Cambridge Univ. Press, Cambridge, 1994).}
\REF\twentyone{M. White, D. Scott and J. Silk, {\it Ann. Rev. Astron.
Astrophys.} {\bf 32}, 319 (1994).}
\refout
\bye